\documentclass[conference,a4paper,10pt]{IEEEtran}

\pdfoutput=1

\usepackage{cite}
\usepackage{graphicx}
\usepackage{tumcolor}

\usepackage{amsmath}
\usepackage{amssymb}
\usepackage{fixmath}
\usepackage[ruled,linesnumbered]{algorithm2e}

\usepackage{color}
\usepackage{bm}

\usepackage[T1]{fontenc}

\usepackage{subfig}

\DeclareMathAlphabet{\mathbit}{OML}{cmr}{bx}{it}

\DeclareMathOperator*{\st}{s.t.\!:}

\DeclareMathOperator{\Hermitian}{H}
\newcommand{\He}{{\Hermitian}}

\DeclareMathOperator{\diag}{diag}

\makeatletter
\let\NAT@parse\undefined
\makeatother

\IEEEoverridecommandlockouts

\begin{document}

\title{User Demand Based Precoding for G.fast Systems}

\author{\IEEEauthorblockN{Andreas Barthelme\IEEEauthorrefmark{1}, Michael Joham\IEEEauthorrefmark{1}, Rainer Strobel\IEEEauthorrefmark{3}, Wolfgang Utschick\IEEEauthorrefmark{1}}
\IEEEauthorblockA{\IEEEauthorrefmark{1}Professur f\"ur Methoden der Signalverarbeitung, Technische Universität München, 80290 München, Germany\\
\{a.barthelme,joham,utschick\}@tum.de}
\IEEEauthorblockA{\IEEEauthorrefmark{3}Intel Connected Home Division, 85579 Neubiberg, Germany, rainer.strobel@intel.com}}

\maketitle

\begin{abstract}
It can be observed that the achievable rate region of a G.fast DSL system is no longer rectangular, as it is the case for vectored VDSL systems, due to stronger crosstalk couplings at high frequencies. Therefore, alternative operating points that are not optimal in a sum-rate sense may be utilized to adapt the system performance to the users' actual demands. To this end, we propose a new precoding scheme based on defining a subset of prioritized users, where we optimize the sum-rate of the prioritized users under a minimum rate guarantee for the remaining users. We present a solution based on Lagrangian duality theory and propose a well-performing one-step heuristic solution. By means of simulations, we show that significant rate gains for the prioritized users can be obtained by the proposed precoding scheme.

\end{abstract}

\IEEEpeerreviewmaketitle

\section{Introduction}
\label{sec:intro}

The fourth generation of copper access technology called G.fast has been standardized by the ITU \cite{G.97012013}. It applies to so called fiber to the distribution point (FTTdp) networks, which are hybrid fiber/copper networks with short copper loop lengths of up to $400$ meters. By increasing the bandwidth by a factor of ten up to $212\,\text{MHz}$, data rates of $1\,\text{Gbit/s}$ and more shall be achieved. With the increasing frequencies, the crosstalk paths become more dominant, which makes interference management essential \cite{Strobel2013}. 

Recent research in the area of precoding for G.fast systems focused on sum-rate optimization. In \cite{Muller2014} and \cite{Timmers2013}, linear and non-linear zero-forcing (ZF) precoding for G.fast systems are presented. Spectrum optimization for linear and non-linear ZF precoding further improves the performance \cite{Strobel2015,Strobel2015b}, which in addition can be enhanced by utilizing generalized inverses \cite{Barthelme2017}. The channel capacity of G.fast systems is investigated via Dirty-Paper-Coding (DPC) in \cite{Lanneer2015}. In \cite{Strobel2015b} and in \cite{Barthelme2016}, linear and non-linear interference allowing precoding techniques based on the weighted minimum mean-squared error (WMMSE) criterion are discussed. It has been shown that the non-linear WMMSE Tomlinson-Harashima-precoding (THP) approaches the sum capacity and that ZF precoding almost achieves the WMMSE THP performance under the typical G.fast scenario \cite{Barthelme2016}. 

With the increased crosstalk at high frequencies, the capacity region of the DSL system changes its shape. Whereas for vectored VDSL systems, the capacity regions featured an almost rectangular structure due to the low crosstalk couplings (see e.g.~\cite{Cioffi2008}), for G.fast systems, the shape of the rate region becomes more radiused. This means that the distance between the sum-rate optimal point (SROP) and the non-achievable utopia point (UP), i.e., the point in which all users operate at their single-user rates simultaneously, increases. In this paper, we change the perspective on the system from a provider-centric point of view, i.e., maximizing the overall throughput, to a user-centric perspective. The increased distance of the SROP to the UP implies for the individual users that the gaps between their performances in the SROP and their single-user performances are big. Hence, the utilization of alternative operating points different from the SROP becomes more interesting for increasing the individual user's experience, by adapting the operating point to the individual data rate demands of the users. While this perspective is common for wireless communication systems (e.g.~\cite{Stanczak2006,Georgiadis2006}), for DSL systems, alternative operating points have not gained a lot of attention since the introduction of vectoring. The idea of adapting the system design to the user demands for DSL systems has been introduced as a part of dynamic spectrum management (DSM) in \cite{Cendrillon2004}, where a power allocation problem with minimum rate constraints is solved. In the context of queue stability, the so called "queue-proportional scheduling" has been proposed for DSL systems with DSM \cite{Cioffi2008}. A more in-depth investigation of cross-layer optimization for DSL systems with DSM can be found in \cite{Tsiaflakis2012}. Recently, van~den~Eynde et al.~proposed a new scheduling algorithm for DSL systems that minimizes the delay violations in order to improve the user experience \cite{vandenEynde2017}. In \cite{Li2012}, Li et al.~combined partial crosstalk cancellation for systems featuring a fixed transmit power allocation with a cross-layer optimization approach to guarantee queue stability and minimize the delay under a fixed budget for the crosstalk canceller taps.

In this paper, we provide a qualitative analysis of the shape of the achievable rate region of G.fast DSL systems. We observe that the achievable rate region is no longer rectangular, as it is the case for vectored VDSL, due to the stronger crosstalk paths (see Section~\ref{sec:rateregion}). This new shape motivates the investigation of alternative operating points. Well known methods to achieve rate points on the boundary are weighted sum-rate optimization and rate balancing, whose objectives are respectively given by
\begin{equation}
\max\,\sum\limits_{\ell=1}^L w_\ell\sum\limits_{n=1}^N r_\ell^{(n)},
\end{equation}
and
\begin{equation}
\max\,\min_\ell\,\frac{\sum\nolimits_{n=1}^N r_\ell^{(n)}}{w_\ell}.
\end{equation}
Unfortunately, the aforementioned approaches cannot guarantee a minimum rate for each user without the knowledge of the shape of the rate region. However, in today's DSL networks, the provider has to guarantee certain rates for different types of plans at all times. Hence, we introduce a new scheduling approach that solves a minimum rate constrained sum-rate maximization problem in Section~\ref{sec:MR}. For the minimum rate constrained problem, we propose a solution based on Lagrangian duality theory and a well performing one-step heuristic solution and compare the methods via simulations. In contrast to the aforementioned existing works on sum-rate suboptimal operating points (e.g. \cite{Cioffi2008,Cendrillon2004,Tsiaflakis2012,vandenEynde2017}), we consider systems with full vectoring and spectrum optimization.

\section{System Model}
\label{sec:systemmodel}
Throughout this paper, we consider the downlink transmission in a multi-user MISO DSL system with $L$ users. Each user is connected to the distribution point (DP) via a twisted pair copper wire, referred to as line. The line lengths in typical G.fast deployments reach up to $400\,\text{m}$. The system uses frequencies from $2\,\text{MHz}$ to $212\,\text{MHz}$. To divide the frequency band into $N=4096$ orthogonal subcarriers, a discrete multitone modulation (DMT) is applied. The uplink and downlink transmission is divided via time division duplexing. We assume that both, transmitter and receivers, have perfect channel knowledge.

\subsection{Linear Precoding}
\label{sec:linprec}
\begin{figure}
	\includegraphics{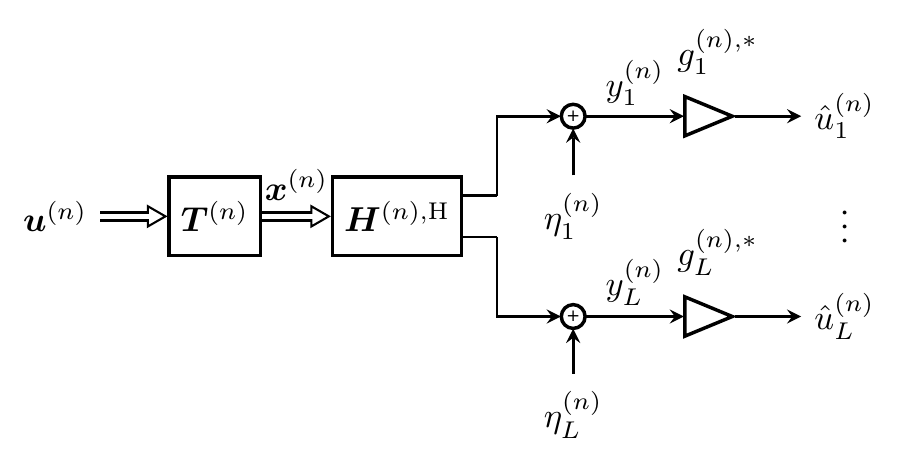}
	\caption{Linear Downlink System Model}
	\label{fig:linsys}
\end{figure}

First, let us consider a system with linear precoding. A block diagram for the corresponding DSL system can be found in Fig.~\ref{fig:linsys}. The modulated data symbols $\bm{u}^{(n)}$ on the $n$-th subcarrier have unit power on average. They are precoded by the linear transmit filter $\bm{T}^{(n)}$ to form the transmit signal $\bm{x}^{(n)}$. The transmit signal is passed through the channel $\bm{H}^{(n)}$ to the non-cooperative receivers. At the receiver of the $\ell$-th user, the signal is perturbed by additive white Gaussian noise $\eta_\ell^{(n)}\sim\mathcal{N}(0,\,\sigma_\ell^{(n),2})$ to form the received signal $y_\ell^{(n)}$. Finally, the received signal is filtered with the scalar receive equalizer $g_\ell^{(n)}$ to form the estimate of the data symbol for user $\ell$ on subcarrier $n$ denoted by $\hat{u}_\ell^{(n)}$.

With this linear system model, the achieved rate of the $\ell$-th user on the $n$-th subcarrier can be expressed as

\begin{equation}
\begin{split}
r_\ell^{(n)}&=\log_2\left(1+\frac{\text{SINR}_\ell}{\Gamma}\right)\\
&=\log_2\left(1+\frac{\|\bm{h}_\ell^{(n),\He}\bm{t}_\ell^{(n)}\|^2}{\Gamma(\sum_{j\neq\ell}\|\bm{h}_\ell^{(n),\He}\bm{t}_j^{(n)}\|^2+\sigma_\ell^{(n),2})}\right),
\end{split}
\end{equation}
where $\text{SINR}_\ell$ stands for signal-to-interference-and-noise ratio of the $\ell$-th user, $\bm{h}_\ell^{(n)}$ and $\bm{t}_\ell^{(n)}$ denote the $\ell$-th channel vector and precoding vector, respectively, and $\Gamma$ is the so-called SNR gap that accounts for losses due to practical coding schemes and the non-Gaussian modulation.

\subsection{Dirty Paper Coding}
Dirty paper coding (DPC) is a non-linear precoding technique that achieves the capacity region of the MIMO broadcast channel \cite{Weingarten2006}. Let us define an encoding order $k_1,\,\dots,\,k_L$. Due to the DPC encoding and decoding operation, the $k_i$-th user sees only the interference from the $k_{i+1}$-th to $k_L$-th user, whereas the interference from the previously encoded users vanishes. This means that the first encoded user sees the whole interference from all other users but the last encoded user sees no interference at all. The DPC encoded data signals propagate over the same transmission chain as seen for the linear case in Fig.~\ref{fig:linsys} with the new special interference properties stemming from the encoding/decoding process. The resulting rate expression for DPC is

\begin{equation}
r_{k_\ell}^{(n)}=\log_2\left(1+\frac{\|\bm{h}_{k_\ell}^{(n),\He}\bm{t}_{k_\ell}^{(n)}\|^2}{\Gamma(\sum_{j>\ell}\|\bm{h}_{k_\ell}^{(n),\He}\bm{t}_{k_j}^{(n)}\|^2+\sigma_{k_\ell}^{(n),2})}\right).
\label{eq:rateexpdpc}
\end{equation}

\subsection{Tomlinson-Harashima Precoding}

\begin{figure*}
	\begin{center}
	\includegraphics{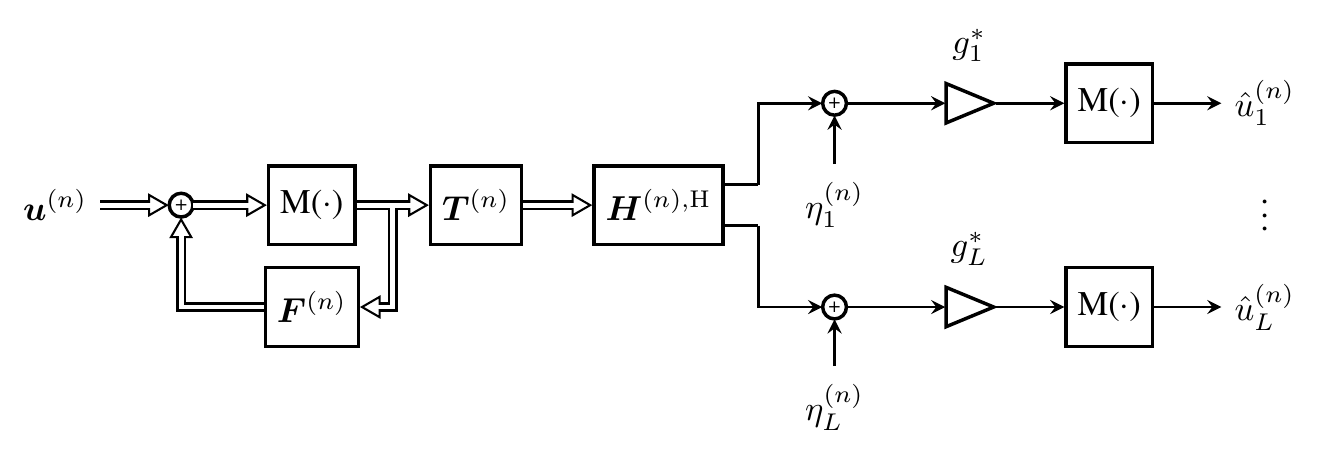}
	\end{center}
	\caption{THP Downlink System Model}
	\label{fig:thpsys}
\end{figure*}
THP is a greedy approach to vector precoding \cite{Schmidt2008}, which approximates the DPC performance \cite{Erez2004}. The system model of a THP system for one subcarrier is depicted in Fig. \ref{fig:thpsys}. The interference cancelation to the previously encoded users, that is typical for DPC, is implemented by a successive feedback loop with an interference canceling feedback filter $\bm{F}$. In the forward path of the feedback loop, a modulo operation is implemented, that greedily chooses the signal point with the lowest power in the corresponding lattice of the used modulation scheme. After the feedback loop, the signal is filtered by another precoding matrix $\bm{T}$ before it is transmitted over the channel. At the receivers, the received signal is filtered by a scalar equalizer $g_\ell$ and is passed through another modulo operation. The modulo operation at the receiver maps the received signal into the Voronoi region of the modulation scheme, which provides the increased degrees of freedom of THP compared to linear precoding \cite{Joham2007}.

After the feedback loop, the signals are approximately uniformly distributed over the Voronoi region. Therefore, an evaluation of the achievable rates is not straightforward. In \cite{Neckebroek2015}, however, a strong relation between the SNR and the achievable bit loadings of ZF THP has been shown. This suggests that a rate evaluation as in (\ref{eq:rateexpdpc}) is quite reasonable for THP.

\subsection{Spectral Constraints}
\label{sec:specconst}
G.fast DSL systems have to fulfill three kinds of constraints. The spectral mask constraint limits the transmit power for each line and subcarrier to a predefined power spectral density (PSD) mask given by the regulation \cite{G.97002013}. These power constraints can be written as the element-wise inequality constraints
\begin{equation}
\diag\left(\bm{T}^{(n)}\bm{T}^{(n),\He}\right)\leq\bm{p}_{\text{mask}}^{(n)}\quad\forall n.
\end{equation}

Second, due to the limited operating range of the transmit amplifiers, the total transmit power per line is limited, i.e.,
\begin{equation}
 \sum\limits_{n=1}^N \diag\left(\bm{T}^{(n)}\bm{T}^{(n),\He}\right)\leq \bm{p}_{\text{sum}}.
 \label{eq:sumpowerconstraint}
\end{equation}

Finally, we want to consider the finite maximal alphabet size of the QAM modulation. For G.fast systems, the highest possible QAM alphabet size is typically for $b_\text{max}=12\,\text{bit}$, with the newly standardized option to extend the alphabet to $14\,\text{bit}$ \cite{G.97012013}. The maximum alphabet size upper bounds the achievable rate, i.e.,

\begin{equation}
r_\text{max}\leq b_\text{max}.
\end{equation}
Whereas for ZF precoding this rate upper bound can be directly transformed into a power constraint \cite{Strobel2015b} due to its interference-free character, for interference allowing precoding methods such as WMMSE precoding or DPC, the maximum bit loading constraint is non-convex and non-increasing in $\bm{T}$. Several interference allowing precoding approaches, however, are based on these properties. In those cases, the maximum bit loading constraints can only be approximated by a tighter constraint that neglects the interference terms (cf.\cite{Barthelme2016}). However, since the maximum bit loading constraints are only active for a very high SNR, and the optimal precoder structure suppresses interference in this region, the losses due to the proposed approximation are negligible.

\section{Rate Region}
\label{sec:rateregion}
Consider the following scenario of two groups of users that differ in their data rate demands for a certain time slot. The first group of users employs application services with a high demand on data rate, such as streaming videos or downloading large files. Naturally, these users want as much available data rate as possible allocated to them at this time slot. On the other hand, there are users that are idle or simply browsing, i.e., these users do not need the full data rate allocated to them as in the SROP. The first set of prioritized users is denoted by $\mathbb{L}_\text{p}$, whereas the set of remaining users is denoted by $\overline{\mathbb{L}}_\text{p}$. We want to answer the question how much could the users in $\mathbb{L}_\text{p}$ profit by limiting the rates of the users in $\overline{\mathbb{L}}_\text{p}$. To find an answer to this question, we refer to the achievable rate region of DSL systems.

It has been shown that for vectored VDSL systems, the capacity region is almost rectangular, i.e., the users can reach their single-user performances simultaneously, e.g., with \cite{Ginis2002,Cendrillon2007}. This result comes from the diagonally dominant structure of the channel matrices of VDSL systems. Employing vectoring, the channel can be diagonalized, while almost keeping the same direct channel gains, with a precoder that has rows with unit norms, therefore, retaining the spectral masks after precoding. A rectangular shaped rate region means, that limiting the rates for $\overline{\mathbb{L}}_\text{p}$ does not enhance the rates of the prioritized users $\mathbb{L}_\text{p}$ as they already achieve their optimal performance in the SROP.

\begin{figure}
	\includegraphics{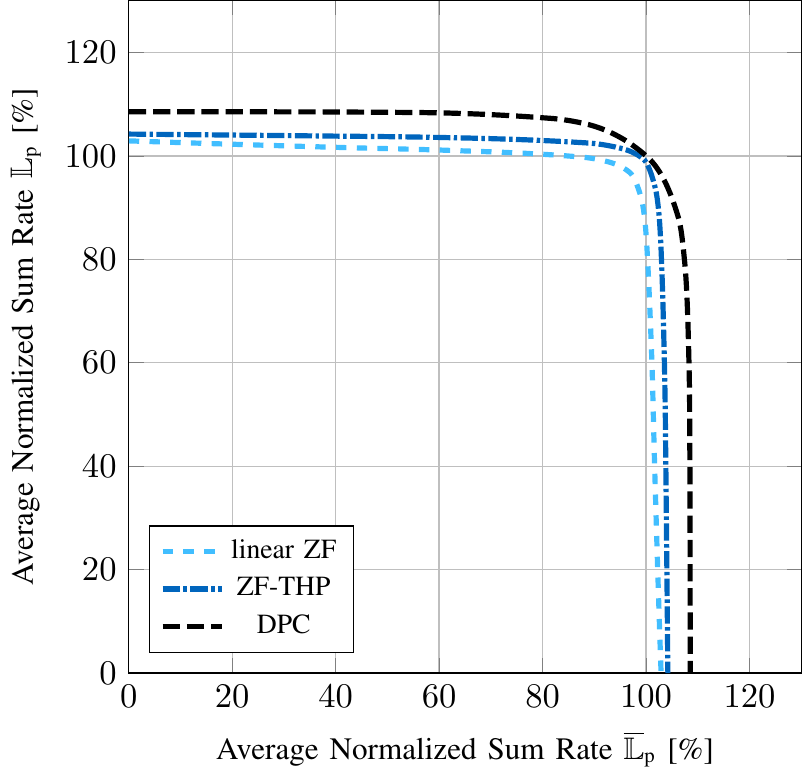}
	\caption{Average Rate Region for G.fast with $|\mathbb{L}_\text{p}|=15$ and $|\overline{\mathbb{L}}_\text{p}|=15$}
	\label{fig:rateregion15}
\end{figure}

\begin{figure}
	\includegraphics{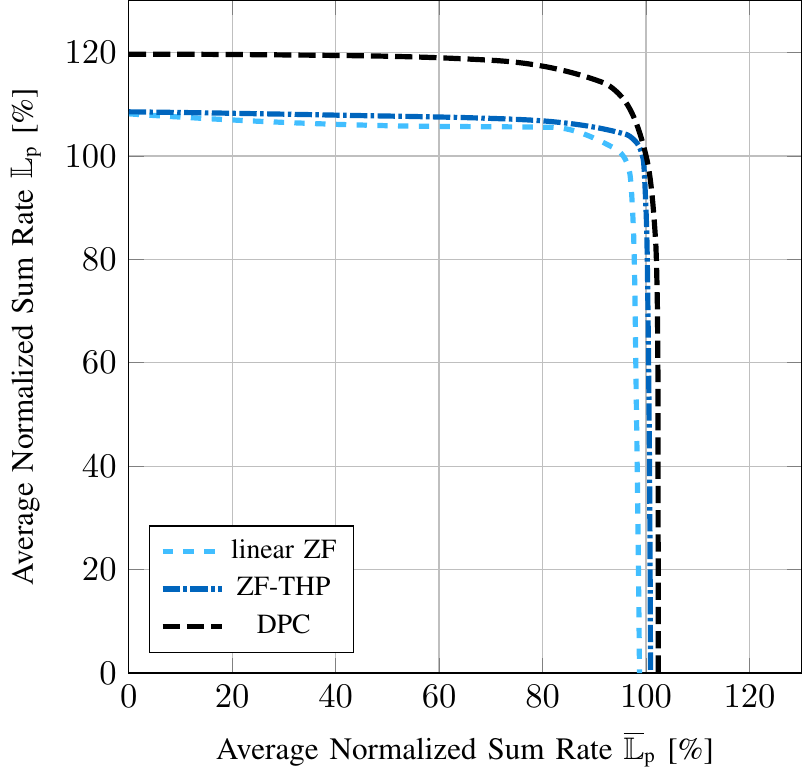}
	\caption{Average Rate Region for G.fast with $|\mathbb{L}_\text{p}|=5$} and $|\overline{\mathbb{L}}_\text{p}|=25$
	\label{fig:rateregion5}
\end{figure}

With the increased crosstalk in G.fast systems, diagonalization without increasing the transmit power is no longer possible, which makes precoding techniques with spectrum optimization, as in \cite{Strobel2015b}, necessary. As a result, the achievable rate region becomes more radiused, i.e., the distance between the SROP and the UP increases. Using DPC, as has been discussed in \cite{Lanneer2015} for G.fast channels, the capacity region can be achieved. Additionally accounting for the maximum bit loading constraints and the SNR gap $\Gamma$ in the DPC scheme, we simulated the achievable rate regions for $10$ cable bundles with $30$ lines each. For this, we solved a weighted sum rate maximization, where we assign the users among each group a common weight. For the linear ZF and ZF THP methods we used the methods discussed in \cite{Strobel2015b}, i.e., we determine the structural part of the precoders to fulfill the ZF constraints, and then, compute the optimal power allocation by solving a convex optimization problem. For THP scheme we applied an encoding order according to the users' weights combined with the "shortest lines last" policy for users with equal weights (for details see Subsection \ref{sec:order}). The results can be found in Figs.~\ref{fig:rateregion15} and~\ref{fig:rateregion5}. 

For the simulations, we plotted the average normalized sum-rate of all prioritized users normalized by the sum of their respective rates in the SROP. Therefore, the SROP is located at $(100\%,100\%)$. Accordingly, the maximal average sum-rate gain of the prioritized users can be found on the ordinate. This is the point where the lines to the users $\overline{\mathbb{L}}_\text{p}$ are only used to improve the transmission of the users $\mathbb{L}_\text{p}$.

The average achievable rate region for two groups with $15$ users each is plotted in Fig.~\ref{fig:rateregion15}. We see that with DPC, an average sum-rate gain of about $10\%$ is possible, if the other $15$ users are used to boost the prioritized users' rates. With a ZF precoding scheme an average sum-rate gain of $5\%$ can be achieved. Gains up to $20\%$ with DPC and $10\%$ with ZF can be achieved for groups of $5$ prioritized users as shown in Fig.~\ref{fig:rateregion5}. Note that the average rate region is pretty flat at the edges, i.e., where the average normalized sum rate of one priority group is below about $80\%$. This is a result of the fact that a large portion of the normalized sum rate of a group stems from the rate of well performing users, whose rate gains are limited because of the maximal alphabet size. This motivates us to take a closer look at the performance of individual users. 

\begin{figure}
	\includegraphics{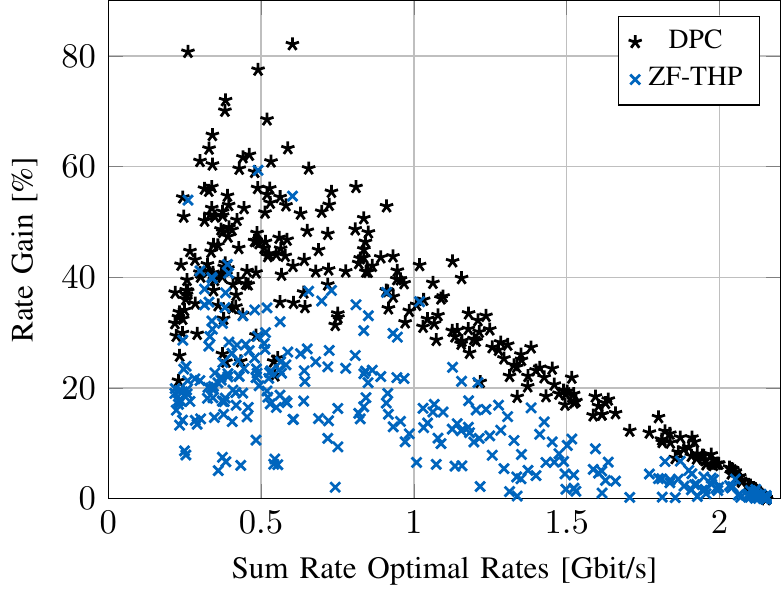}
	\caption{Maximal Individual Rate Gains vs Achieved Rates in SROP for $|\mathbb{L}_\text{p}|=5$ and $|\overline{\mathbb{L}}_\text{p}|=25$} when the Users in $\overline{\mathbb{L}}_\text{p}$ are not served.
	\label{fig:indrategains5}
\end{figure}

\begin{figure}
	\includegraphics{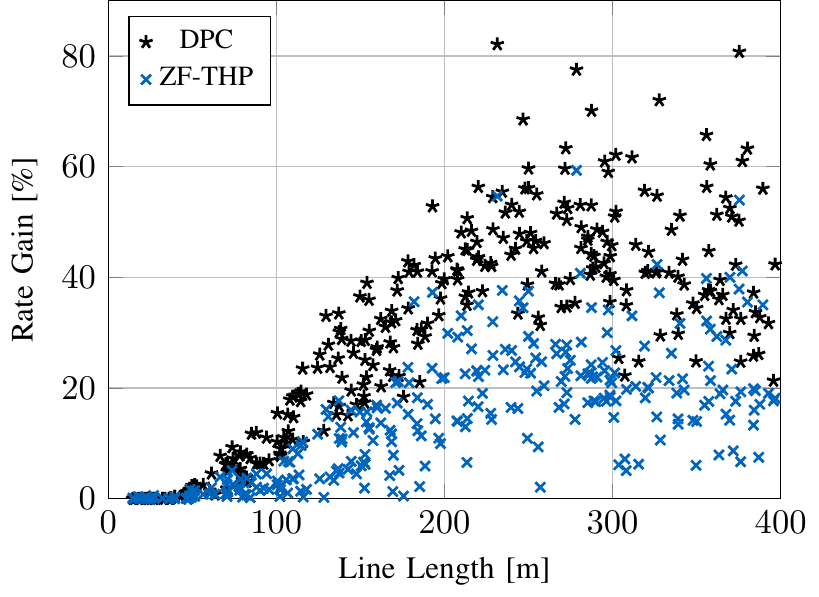}
	\caption{Maximal Individual Rate Gains vs Line Length for $|\mathbb{L}_\text{p}|=5$ and $|\overline{\mathbb{L}}_\text{p}|=25$} when the Users in $\overline{\mathbb{L}}_\text{p}$ are not served.
	\label{fig:indrategains5length} 
\end{figure}

 Now, we consider the individual rate gains of the prioritized users in the extreme case where the users in $\overline{\mathbb{L}}_\text{p}$ are only used to boost the prioritized users' rates, i.e., that no data is intended for the users in $\overline{\mathbb{L}}_\text{p}$. For the case of $5$ prioritized users, we plotted the rate gains compared to the users' performances in the SROP in Fig.~\ref{fig:indrategains5}. For a setup with $|\mathbb{L}_\text{p}|=5$, we obtain $5$ points in Fig.~\ref{fig:indrategains5} for a certain channel realization and prioritized user selection, i.e., one for each prioritized user. Since we employ a round robin user selection, such that each user is prioritized exactly once, each channel realization generates $30$ points in Fig.~\ref{fig:indrategains5}, leading to $300$ points for $10$ channel realizations. To keep the performance of DPC and ZF methods still distinguishable, the results for linear ZF have been omitted, as the results are very similar to the ZF THP results. We see that for individual users, gains of $80\%$ for DPC and $60\%$ for ZF THP are achievable. The highest gains are obtained for the low performing users, whereas the gains for the already well performing users are limited by the finite QAM alphabet size. The low performing users are the users that suffer the most from crosstalk from other lines, which explains the high potential gains for these users. In Fig.~\ref{fig:indrategains5length}, the individual rate gains have been plotted against their respective line length. This figure reflects the correlation between line length and performance in the SROP, i.e., the users with long lines can potentially benefit a lot more from prioritizing than the ones with short lines.

\section{Minimum Rate Constrained Sum Rate Maximization}
\label{sec:MR}
Finding a suitable model for the internet traffic has proven to be difficult, however, several publications, for example \cite{Hernandez2004}, suggest that the internet traffic features a heavy-tailed, long range-dependent distribution. Classic scheduling approaches such as the maximum weight scheduling (MWS), which is a weighted sum-rate optimization approach where the weights depend on the individual users queue fill levels, are not inherently suited to cope with these kind of traffic properties. The bursty arrival processes lead to a poor delay performance of MWS, as has been shown in \cite{Markakis2009,Jagannathan2013}. Instead of augmenting the MWS algorithm to cope with the heavy tailed distribution as discussed in \cite{Markakis2009,Jagannathan2013}, we want to take a different approach. We propose to split the users in two priority groups according to their data rate demands as described in Section \ref{sec:rateregion}. The users of group $\mathbb{L}_\text{p}$ get the maximally possible rates under the constraint that the users of group $\overline{\mathbb{L}}_\text{p}$ have a minimum rate guaranteed. Thereby, we assume that the assignment to each group has been determined by an unspecified outer algorithm that may be based on, e.g., DSL plans with differing QoS guarantees for different users, aggregated queue-levels, and/or user behavior predicted by machine learning algorithms.

This priority group based approach has two advantages. First, by guaranteeing a minimum rate for all users, the scheme is more marketable as a service, as the guaranteed rate ensures the functionality of a base set of applications at all times. On the contrary, MWS or rate balancing cannot inherently guarantee a minimum rate for users due to the lack of knowledge of the shape of the rate region. The second advantage of the proposed scheduling scheme is the relaxed time constraint for precoder computations. For queue level based scheduling approaches, the precoder has to be determined for every change in the queue levels to track the user demands. For the proposed scheme, the users of $\mathbb{L}_\text{p}$ operate at their maximum achievable rate, and the other users at the guaranteed rate. It is crucial, that the minimum rate is chosen such that the bursty transmissions occurring by simple browsing can be easily coped with. Hence, the precoders have to be recalculated only if a user switches from one group to another. 

\subsection{Algorithm}
The proposed scheme can be cast as a minimum rate constrained sum-rate optimization problem of the following form

\begin{equation}
\begin{split}
	&\max\limits_{\bm{T}\in\mathcal{S}}\,\sum\limits_{n=1}^N \sum\limits_{\ell\in\mathbb{L}_\text{p}} r_\ell^{(n)}\\
	\st &\sum\limits_{n=1}^N r_\ell^{(n)}\geq r_{\text{min},\ell} \quad \forall \ell\in\overline{\mathbb{L}}_\text{p},
\end{split}
\label{eq:minrateprob}
\end{equation}
where the set $\mathcal{S}$ accounts for the power constraints presented in Subsection \ref{sec:specconst}.

Note that we restrict ourselves to choose $r_{\text{min},\ell}$ to be smaller than the $\ell$-th user's performance in the SROP, as we want to consider the scenario described above, where the resources of low demanding users are used to improve the performance of users with high rate demands. Otherwise, the optimization problem (\ref{eq:minrateprob}) can be used for the reach enhancement of a given service (cf. \cite{Strobel2017}).

\subsubsection{Alternating Approach}
\label{sec:MRaa}
Assuming ZF precoding, the optimization in (\ref{eq:minrateprob}) is convex, since the resulting rate expressions are concave functions of the power allocation (cf. \cite{Strobel2015}). We can employ Lagrangian duality theory and dualize the minimum rate constraints to obtain
\begin{equation}
	\min_{\bm{0}\leq\bm{\lambda}}\,\max\limits_{\bm{T}\in\mathcal{S}}\,\sum\limits_{n=1}^N \sum\limits_{\ell\in\mathbb{L}_\text{p}} r_\ell^{(n)}+\sum\limits_{\ell\in\overline{\mathbb{L}}_\text{p}}\lambda_\ell\left(\sum\limits_{n=1}^N r_\ell^{(n)}- r_{\text{min},\ell}\right).
	\label{eq:dualprob}
\end{equation}

The inner maximization can be easily identified as a weighted sum-rate optimization problem, where the users of $\mathbb{L}_\text{p}$ are weighted with $1$, and the Lagrange multipliers $\bm{\lambda}$ define the weights for the users of $\overline{\mathbb{L}}_\text{p}$. 

As a weighted sum-rate optimization problem, the inner problem can be solved by well-known algorithms as presented in \cite{Strobel2015,Strobel2015b}. For the outer optimization, we propose a subgradient projection update
\begin{equation}
	\lambda_\ell^{[t+1]}=\max\{0,\,\lambda_\ell^{[t]}+\alpha^{[t]}(r_{\text{min},\ell}-\sum\limits_{n=1}^N r_\ell^{(n)})\}\,\forall\ell\in\overline{\mathbb{L}}_\text{p},
	\label{eq:subgrad}
\end{equation}
where $t$ denotes the iteration index and $\alpha^{[t]}$ is an adaptive step size for the $t$-th iteration. 

For interference allowing precoding methods, the minimum rate constrained sum-rate optimization problem is non-convex due to the interference terms in the denominators of the SINR expressions. However, as has been shown in \cite{Yu2006b}, strong duality holds even for non-convex optimization problems as long as the so called time-sharing property holds which is usually --- at least approximately --- the case for multi-carrier systems. In fact, the optimization problem in (\ref{eq:minrateprob}) fulfills the time-sharing property, and we can thus dualize the minimum rate constraints to obtain the dual problem (\ref{eq:dualprob}).

This means that to solve this minimum rate constrained sum-rate maximization, a weighted sum-rate maximization problem has to be solved $N_\text{it}$ times. Here, $N_\text{it}$ denotes the number of outer optimization steps, i.e., the number of subgradient projection steps (\ref{eq:subgrad}) needed for the Lagrange multipliers to converge. This number naturally depends on the channel realization, initialization of $\bm{\lambda}$, and the step size control.

Note that as we solve a weighted sum rate optimization in (\ref{eq:dualprob}), in theory, the achieved rates for some users of $\mathbb{L}_\text{p}$ may actually deteriorate. However, in our considered scenario with $r_{\text{min},\ell}$ smaller than the performance in the SROP, we did not observe this behavior in our simulations. To guarantee that this effect does not occur, minimum rate constraints for the prioritized users can be added to (\ref{eq:dualprob}) that guarantee the same performance as in the SROP.

It has been shown in \cite{Strobel2015b} that for making ZF precoding techniques for DSL systems competitively viable, a smart heuristic of disabling the subcarrier-line combinations which do not achieve at least $1$ bit, i.e., adding these subcarrier-line combinations to the set of inactive subcarrier-line combinations $\mathbb{I}_{\text{dis}}$, is very effective. As disabling a subcarrier-line combination may improve the performance of another, it is important that we update the set $\mathbb{I}_{\text{dis}}$ successively. In our simulations, we also observed that disabling the weakest user performing under $1$ bit on each subcarrier in parallel performs almost as well as iteratively disabling the overall weakest subcarrier-line combination, while reducing the computational complexity. To combine this heuristic with the optimization (\ref{eq:minrateprob}), we investigated an alternating approach, where we solve (\ref{eq:minrateprob}) in one step for fixed $\mathbb{I}_\text{dis}$, and then updating the set $\mathbb{I}_\text{dis}$ heuristically in the next step. In the case of the minimum rate constrained sum-rate optimization, it turned out to be effective to disable the worst performing subcarrier-line combination of the users without prioritization where rates below $1\,\text{bit}$ are achieved before disabling low performing subcarriers of users in $\mathbb{L}_\text{p}$. 

\subsubsection{One-Step Heuristic Solution}
\label{sec:MRheu}
\begin{figure}
	\centering
		\subfloat[Average Rate Per Subcarrier of Users in $\mathbb{L}_\text{p}$]{
		\includegraphics{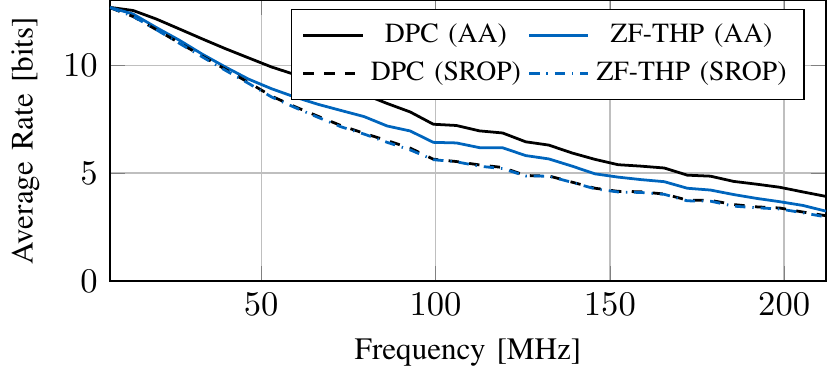}
		}\newline
	\subfloat[Average Rate Per Subcarrier of Users in $\overline{\mathbb{L}}_\text{p}$]{
		\includegraphics{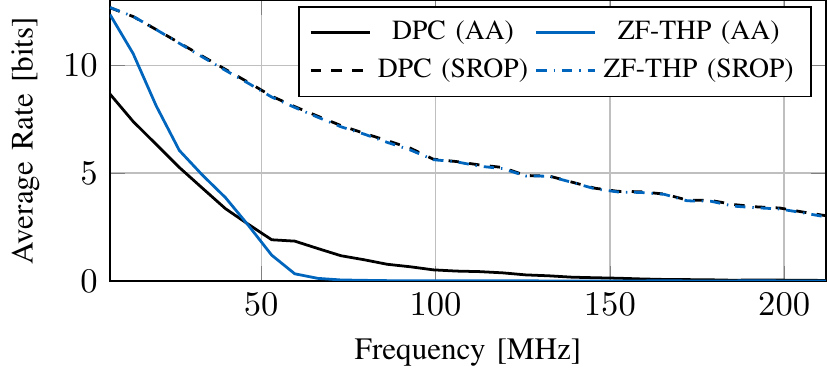}
}\newline
\subfloat[Amount of Disabled Lines Per Carrier]{
	\includegraphics{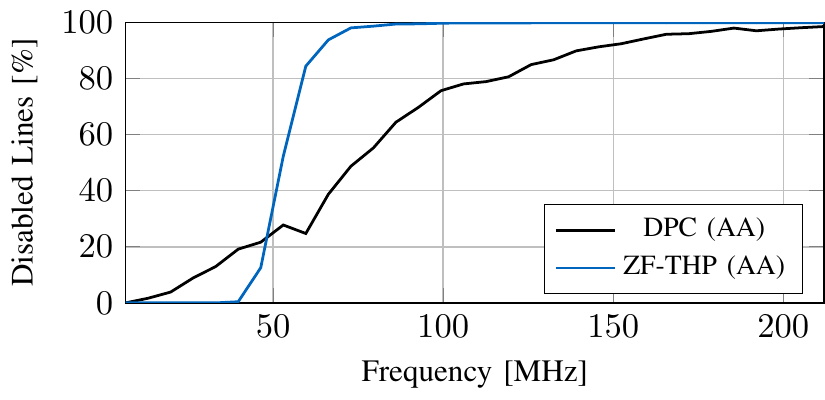}
}
	\caption{Statistics for the Alternating Approach (AA).}
	\label{fig:probdisabled}
\end{figure}

Now, we present an one-step heuristic solution to the minimum rate constrained sum-rate maximization (\ref{eq:minrateprob}) to reduce the computational complexity of the previously described approach. It can be observed, that due to the special channel properties of DSL cables, for the minimum rate constrained users, subcarrier-line combinations at higher frequencies will be disabled more frequently by the alternating approach (see Figure \ref{fig:probdisabled}). The figure shows that especially for ZF precoding, the high frequencies of the users in $\overline{\mathbb{L}}_\text{p}$ are only used to improve the performance of the prioritized users $\mathbb{L}_\text{p}$. For DPC, the used frequency range is a bit broader, which can be explained by its inherent ability to improve the prioritized users' rates by its optimal interference management without the need to disable subcarrier-line combinations to achieve this additional degree of freedom. DPC is able to optimally distribute resources between two users because it can allow more or less interference to a certain user; for ZF, the decision is more binary as it either allows interference to a certain user, if the user is disabled, or completely cancels the interference to any active user.

This leads us to the following heuristic. Based on the sum-rate optimal solution with its rate allocation $\{r_{\text{sum},\ell}^{(n)}\}$, we allocate as many subcarriers to each minimum rate constrained line as needed to fulfill the rate lower bound, starting at the lowest frequency. The allocation strategy is summarized in Algorithm~\ref{alg:heu}.
\begin{algorithm}
	\DontPrintSemicolon
	$n_{\text{max},\ell}=0\quad\forall\ell\in\overline{\mathbb{L}}_\text{p}$\;
	\For{$\forall\ell\in\overline{\mathbb{L}}_\textnormal{p}$}{
	\Repeat{$\sum_{n=1}^{n_{\textnormal{max},\ell}}r_{\textnormal{sum},\ell}^{(n)}\geq r_{\textnormal{min},\ell}$}{$n_{\textnormal{max},\ell}=n_{\textnormal{max},\ell}+1$\;}
	Add $(n,\ell)$ to $\mathbb{I}_\textnormal{dis}$ for all $n>n_\textnormal{max}$\;}
	\caption{Heuristic Subcarrier Allocation}
	\label{alg:heu}
\end{algorithm}

For each user, we disable all the subcarriers that have not been allocated by Algorithm \ref{alg:heu}, i.e., we no longer transmit data to user $\ell$ over all subcarriers $n>n_{\text{max},\ell}$. However, the disabled lines per subcarrier are still used to transmit data to all the active users on the respective subcarrier. To achieve the gain of disabling certain lines on certain subcarriers, the precoders have to be recomputed. To this end, we maximize the sum-rate of all users with the new set of disabled subcarrier-line combinations $\mathbb{I}_\text{dis}$ as in \cite{Strobel2015,Strobel2015b,Barthelme2016}, i.e., by applying the respective pseudo-inverses of the resulting non-square channels for the ZF approaches and forcing the respective power values to zero for DPC precoder optimization.

Theoretically, for ZF precoding, the solution to the sum-rate problem for the updated $\mathbb{I}_\text{dis}$ could deteriorate the aggregated rate of a user in $\overline{\mathbb{L}_\text{p}}$, whenever it is better to redistribute power from an active to a disabled subcarrier. Neglecting the spectral mask constraints, and assuming that the power-allocation per user and subcarrier roughly follows a water-filling scheme, this would always be the case if the water-level after updating $\mathbb{I}_\text{dis}$ decreases, i.e., whenever the sum of the reciprocals of the direct channel gains at the disabled subcarriers is smaller than that of the crosstalk paths. However, for any $r_{\text{min},\ell}$ that is substantially smaller than the performance of user $\ell$ in the SROP, this effect is highly improbable for typical G.fast channels, and has never occurred in our simulations.

Obviously, the proposed heuristic significantly reduces the computational complexity of solving (\ref{eq:minrateprob}). Where we needed to solve a weighted sum-rate maximization problem $N_\text{it}$ times for each update of $\mathbb{I}_\text{dis}$ in the alternating approach, we now compute the precoders exactly once.

\subsection{Ordering}
\label{sec:order}
In THP systems, the choice of the encoding order has a significant impact on the performance. Optimizing the encoding order with respect to some objective is in general non-trivial and leads to a combinatorial search. For sum-rate optimization, the almost optimal encoding order follows the "best last" policy \cite{Kusume2007}. For G.fast systems, this encoding strategy can be translated to a "shortest lines last" policy, as the cable attenuation heavily correlates with the channel quality in most cases.

Considering fairness criteria, the optimization of the encoding order has been investigated in the context of G.fast DSL systems in \cite{Hekrdla2015,Zu2016,Lanneer2017b}.

For fixed $\mathbb{I}_\text{dis}$, the minimum rate constrained sum-rate optimization can be solved by a weighted sum-rate optimization problem, where the weights are given by the Lagrange multipliers of the QoS constraints. The optimal encoding order for such a problem is given by the users' weights in descending order. Unfortunately, the optimal Lagrange multipliers itself depend again on the encoding order. Therefore, an iterative update of encoding order and optimal Lagrange multipliers may not converge. Hence, we propose a suboptimal heuristic encoding order based on the partition into priority groups. As we want to maximize the sum-rate of the users in $\mathbb{L}_\text{p}$, we propose to encode these users first, i.e., during the encoding process, these users do not consider the interference to the remaining users in $\overline{\mathbb{L}}_\text{p}$. Among users in one of the priority groups $\mathbb{L}_\text{p}$ and $\overline{\mathbb{L}}_\text{p}$, we employ the "shortest lines last" policy described above. This means that for fixed priority groups, the encoding order is fixed.

\section{Simulation Results}
\label{sec:simres}
To compare the performance of the algorithms presented in the previous section, we ran several simulations for G.fast channels generated according to the channel models presented in \cite{Strobel2013}. We consider deployments with $30$ lines per bundle, where the lines range from $10$ to $400\,\text{m}$. The used frequency band ranges from $2$ to $212\,\text{MHz}$; accordingly, a DMT of size $4096$ is applied. Our algorithms account for the maximum QAM alphabet size of $12\,\text{bit}$. We assume a flat noise power spectral density of $-140\,\text{dBm/Hz}$ on each line. According to the ITU-R standard \cite{G.97002013}, the sum power per line is limited by $4\,\text{dBm}$ and the spectral mask ranges from $-65\,\text{dBm/Hz}$ to $-79\,\text{dBm/Hz}$. Whenever results for THP are shown, the THP specific losses in the form of power loss, modulo loss, and shaping loss are neglected. For each channel realization, we chose the prioritized users in a round robin fashion, ensuring that in each channel realization every user is prioritized exactly once. For the precoder optimization, we followed the lines of \cite{Strobel2015b} for ZF and \cite{Lanneer2015} for DPC, as already discussed in Section~\ref{sec:rateregion}.

\begin{figure*}
	\includegraphics{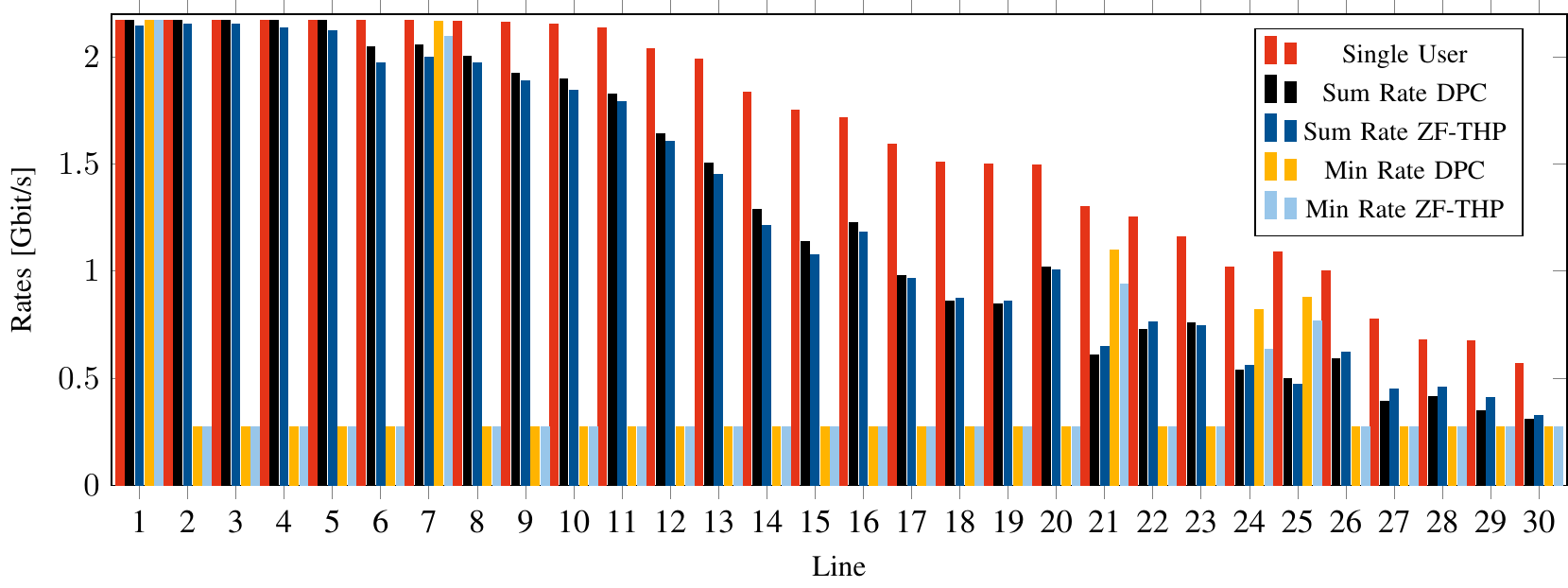}
	\caption{Exemplary Binder with $|\mathbb{L}_\text{p}|=5$ and $r_\text{min}=250\,\text{Mbit/s}$}
	\label{fig:exampleminrate}
\end{figure*}

\begin{figure}
	\includegraphics{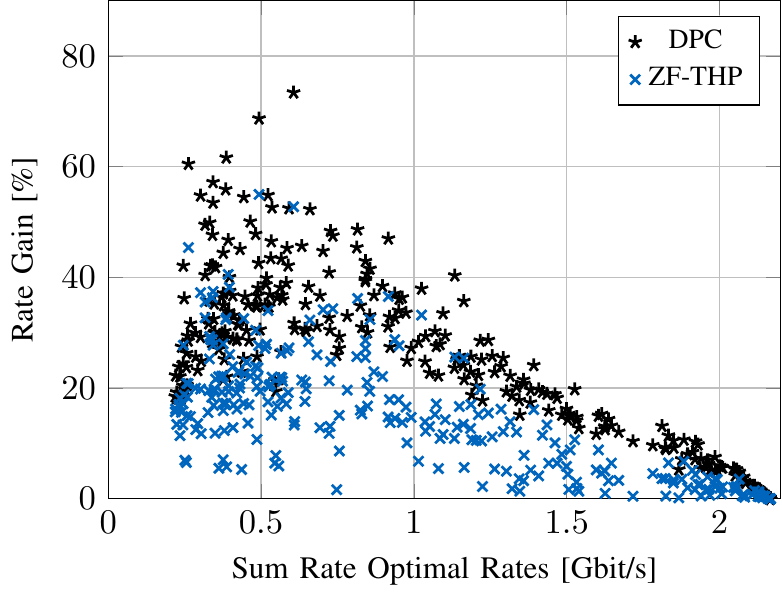}
	\caption{Individual Rate Gains for $|\mathbb{L}_\text{p}|=5$ and $r_{\text{min}}=250\,\text{Mbit/s}$ from Heuristic Optimization}
	\label{fig:minrategains}
\end{figure}

Let us consider a scenario with $|\mathbb{L}_\text{p}|=5$ prioritized users, where the remaining $|\overline{\mathbb{L}}_\text{p}|=25$ users fall back to a minimum rate of $250\,\text{Mbit/s}$. First, we present the results for the heuristic optimization approach to the minimum rate constrained sum-rate maximization. For a better understanding, we illustrated the results for one channel realization in Fig.~\ref{fig:exampleminrate}. In this example, the users $1$, $7$, $21$, $24$, and $25$ have been prioritized. As expected, the minimum rate constrained sum-rate maximization improves the performance of the prioritized users compared to their performance in the SROP. Additionally, we can identify the gap between the sum-rate optimal performance, and the individiual users' single-user rates discussed in Section \ref{sec:rateregion}. Similar to Fig.~\ref{fig:indrategains5}, we look at the individual rate gains of the prioritized users with respect to their performance in the SROP (see Fig.~\ref{fig:minrategains}). Again, we see that the achievable gains for the already well performing users are limited by the QAM alphabet size. However, for the users that are low performing in the SROP, individual gains up to $75\%$ for DPC and about $55\%$ for ZF THP are possible. 

\begin{figure}
	\includegraphics{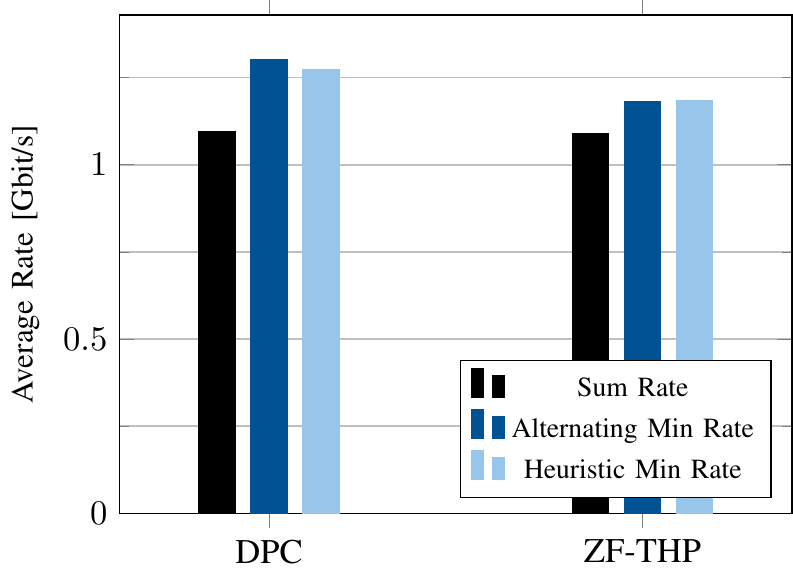}
	\caption{Average Normalized Rate of Prioritized Users for $|\mathbb{L}_\text{p}|=5$ and $r_\text{min}=250\,\text{Mbit/s}$}
	\label{fig:avgnormrate}
\end{figure}

Now, we compare the two different solution approaches to the minimum rate constrained sum-rate maximization for the same scenario as considered above. In Fig.~\ref{fig:avgnormrate}, we show the individual rates of the prioritized users averaged over the prioritized user selection and channel realizations for both DPC and ZF THP. On average, an individual rate gain of about $20\%$ is possible. For the DPC solution, the alternating solution (see Subsection \ref{sec:MRaa}) performs better than the proposed one-step heuristic (see Subsection \ref{sec:MRheu}). On the other hand, for ZF precoding, the alternating approach is slightly worse than the one-step heuristic solution approach. We think that the difference stems from the different interference management of ZF and DPC. The DPC algorithm can already exploit interference to improve certain lines at the cost of unprioritized lines, whereas the ZF algorithms have no inherent knowledge on the effect of disabling certain subcarrier-line combinations. This may lead to a better choice for disabling subcarrier-line combinations in the case of DPC compared to ZF methods.

\section{Conclusion}
This paper adopts a new perspective on precoding for G.fast DSL systems by considering the individual users' demands instead of optimizing the overall throughput of the system. As we have shown, this point of view is motivated by the shape of the achievable rate regions of G.fast systems. Due to the increased crosstalk, the rate regions no longer feature a rectangular structure as it is the case for vectored VDSL systems. Utilizing sum-rate suboptimal operating points, the individual users rates can be significantly increased. In this paper, we have presented a special method for obtaining such alternative operating points based on a minimum rate constrained optimization. The resulting optimization problem can be solved by means of Lagrange duality and leads to a two step optimization, where we have an inner weighted sum-rate maximization problem and an outer optimization over the Lagrange multipliers that determines the optimal weights for the respective objective. Additionally, we proposed an one-step heuristic solution that decreases the computational complexity significantly. By means of simulations, we have shown that the minimum rate constrained sum-rate optimization can improve the individual users' performances by up to $75\%$ percent compared to the performance in the SROP. Especially for the low performing longer lines, this sum-rate suboptimal approach may increase the individual users' rates significantly, whereas the already well performing users' gains are limited by the finite QAM alphabet size.

\section{Future Work}
\begin{figure}
	\includegraphics{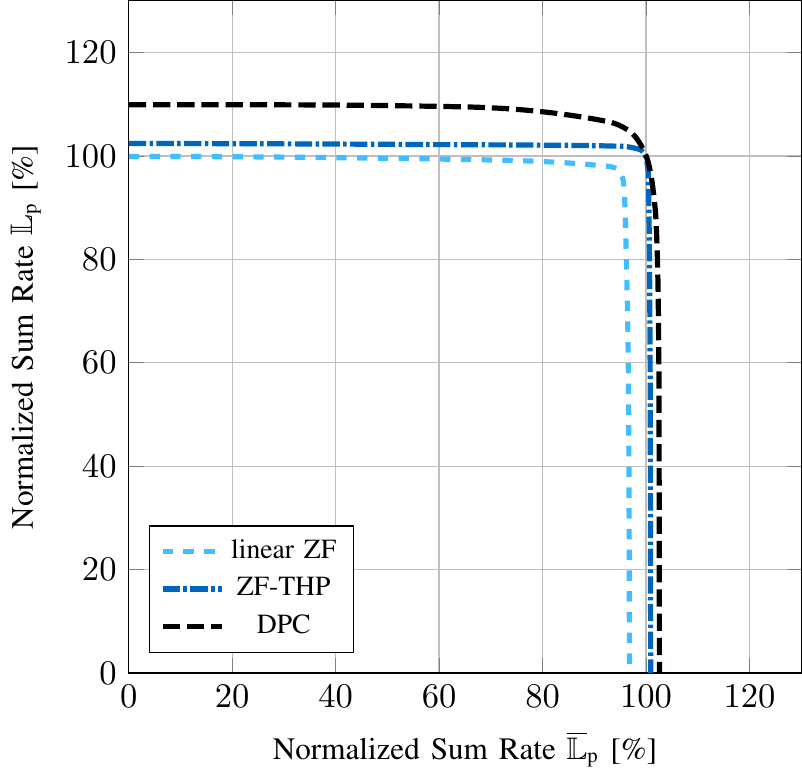}
	\caption{Rate Region for G.mgfast cable U72 $|\mathbb{L}_\text{p}|=2$ and $|\overline{\mathbb{L}}_\text{p}|=6$}
	\label{fig:rateregionM}
\end{figure}

With the expansion of the frequency band towards $1\,\text{GHz}$, as currently discussed for the G.fast successors, the crosstalk paths will get even more dominant, which makes the utilization of alternative operating points even more appealing. On the other hand, the line lengths and the number of lines per binder decrease, which may limit potential gains. As a preliminary assessment of the potential of alternative operating points, we determined the rate region for a Swisscom measurement of an U72 cable with eight lines (see Fig.~\ref{fig:rateregionM}) \cite{Swisscom17}. The simulation parameters for the G.mgfast system are similar to the parameters used for the G.fast simulation that has been presented in Section \ref{sec:simres} except that the G.mgfast system uses frequencies up to $848\,\text{MHz}$ and the number of prioritized users is $|\mathbb{L}_\text{p}|=2$. In the exemplary binder, the line length for each user is $25\,\text{m}$.

From Fig.~\ref{fig:rateregionM}, we see that the rate region of the exemplary binder is again non-rectangular. This motivates further research on the achievable rate regions for different cable types and line lengths of G.mgfast systems.

Apart from that, our results in Fig.~\ref{fig:rateregionM} and Section~\ref{sec:rateregion} and \ref{sec:simres} have shown that the ZF approaches are no longer able to attain the DPC performance highly prioritized scenarios contrary to results for the SROP. Therefore, interference allowing precoding techniques should be reevaluated in this regard. Additionally, sum-rate suboptimal operating points may be utilized for other than the considered objectives, e.g., delay optimization. 
\bibliographystyle{IEEEtran}
\bibliography{UserDemandBasedPrecodingGfast}

\end{document}